# A Novel IoT Architecture based on 5G-IoT and Next Generation Technologies

Hamed Rahimi, Ali Zibaeenejad, Ali Akbar Safavi
Department of Electrical and Computer Engineering, Shiraz University, Shiraz, Iran
Email: {hamed.rahimi, zibaeenejad, safavi}@shirazu.ac.ir

*Abstract*— The Internet of Things (IoT) is a crucial component of Industry 4.0. Due to growing demands of customers, the current IoT architecture will not be reliable and responsive for next generation IoT applications and upcoming services. In this paper, the next generation IoT architecture based on new technologies is proposed in which the requirements of future applications, services, and generated data are addressed. Particularly, this architecture consists of Nano-chip, millimeter Wave (mmWave), Heterogeneous Networks (HetNet), device-to-device (D2D) communication, 5G-IoT, Machine-Type Communication (MTC), Wireless Network Function virtualization (WNFV), Wireless Software Defined Networks (WSDN), Advanced Spectrum Sharing and Interference Management (Advanced SSIM), Mobile Edge Computing (MEC), Mobile Cloud Computing (MCC), Data Analytics and Big Data. This combination of technologies is able to satisfy requirements of new applications. The proposed novel architecture is modular, efficient, agile, scalable, simple, and it is able to satisfy the high amount of data and application demands.

## I. INTRODUCTION

the Industry 4.0 represents the Fourth Industrial Revolution (4IR) in automation and data exchange [1]. It refers to several fundamental new technology innovations such as Cyber-Physical Systems (CPS) [2], Internet of Things (IoT) [3], and Data Analytics [4]. Industry 4.0 organizes special connections between physical devices, information, humans, and biological spheres to modify traditional lifestyles [5]. The IoT as one of the most important parts of industry 4.0 becomes extensively demanding and it will probably alter social relations [6]. According to the Statista report, the number of connected devices installed base worldwide in 2025 will be increased to more than 75 billion devices. Telefonica estimated that more than 90 percent of cars by 2020 will be connected to the Internet networks [7]. In near future, the number of new IoT applications and their generated data will be extremely increased. Therefore, the amount of data and customer demands will highly grow up. This circumstance will make a challenging situation because applications will need faster, smarter, simpler, more reliable, and more scalable architecture than before. In fact, the current IoT architecture will not be reliable and responsive for next-generation IoT applications and upcoming services. Hence, it will no longer able to support the IoT devices due to the extremely large number of devices and service requests. The technologies, which have been employed to design the current architectures, will not provide smooth connectivity to a large number of things due to high service requests and data exchange rates. Recently, the communication section of the IoT systems has been considerably developed in such a way that things and devices are connected with various communication protocols such as ZigBee, Bluetooth Low Energy (BLE), Wi-Fi, GSM, Lora, and Sigfox [8]. Consequently, imbalance of the communication protocols can make serious problems such as radio-frequency interference (RFI) [10]. Hence, a novel architecture based on new technologies is required to treat future challenges such as RFI [9]. Also, networks, processes, applications, and data management architectures require a different communication and processing model. Consequently, many researchers are working on this direction [10].

This article is motivated by the future requirements of network architectures to entertain billions of IoT devices. In this paper, we design an eight-layer architecture, which benefits from new technologies such as 5G. The number of layers is selected according to the fact that most of these technologies have separate functionality and thus we embed them in individual layers. This proposed architecture makes analysis, modularity, and scalability of an IoT system more efficient. This model simplifies, clarifies, identifies, standardizes, and organizes the essential component of future IoT systems.

This paper is organized as follows. In Section II, we review several available remarkable architectures in the literature. In Section III, we introduce new technologies, which will develop the main part of IoT systems in near future. In Section IV, we propose a novel IoT architecture based on 5G-IoT and next-generation technologies illustrated in Section III. In Section V, we compare our proposed architecture with the architectures mentioned in Section II. In this section, we establish how the future requirements are met. Finally, we conclude this paper in Section VI.

## II. LITERATURE REVIEW

In this section, several existing IoT architectures are explained. These architectures will be referred in next sections where we establish our new architecture.

**1. Three-Level Architecture.** The three-level architecture is elementary for the IoT, which has been designed and implemented in a number of systems. A general IoT architecture includes three levels: sensing, transport, and application. In [11], a self-configuring and scalable architecture for a large-scale IoT network has been proposed. The IoT can be identified in three levels: Internet-oriented, sensors and actuators, and knowledge. Commonly, it accomplished that the implementation of IoT technology is nearby to modern civilization, where devices and society are united practically to information systems via wireless sensors [12]. Commonly, the architecture of IoT is classified into three primary levels [13]: 1- perception level, 2- network level, and 3- application level. The sensor layer, which also known as Perception level, is implemented as the bottom layer in IoT architecture. The transmission, which also known as Network level, is implemented as the middle layer in IoT architecture [14]. The business layer, which also known as the application layer, is implemented as the top layer in IoT architecture [8]. Examples include smart cities, smart transportation, smart grid, etc.

**2. SDN-Based Architecture.** An SDN-based architecture for the IoT to provide high-level quality of service (QoS) to the various IoT tasks in heterogeneous wireless network environments, is designed by Z. Qin et al. [15].

**3. QoS-Based Architecture.** In [16] Jin et al. propose four different IoT architectures, which allow various smart city applications to include their QoS requirements. The suggested network architectures are 1- autonomous, which bolster Internet disconnected networks; 2- ubiquitous, where Smart Things Networks (STN) are a part of the Internet; 3- application-level overlay, that uses NFV to reduce the stress and congestion among nodes[17]; 4- service-oriented, where specific gateways interacted with the inherent heterogeneity of the IoT environment.

**4. SoA-Based Architecture.** Typically, SoA is a component-based model, which can be designed to connect various functional sections (also known as services) of an application through protocols and interfaces [18], [3]. SoA concentrates on designing the process of coordinated services, and optimize software and hardware segments, increasing the probability of SoA for use in designing IoT architecture[18], [19]. SoA-based IoT architecture consists of four levels which cooperate with each other: 1- the perception level 2- network level 3- service level 4- application layer. In the four-level SoA-based IoT architecture, the perception level is performed as the bottom layer of the architecture and used to sense, storage, and analyze the data associated with physical devices.

**5. MobilityFirst Architecture.** In [20], it was shown that future Internet architecture (FIA) called MobilityFirst by using name-based, could address many challenges associated with Smartphones when acting as casual gateways of WSANs in IoT systems.

**6. CloudThings Architecture.** An IoT-enabled smart home scenario was presented by J. Zhou et al. [21] to analyze the IoT application requirements. So that CloudThings architecture has been proposed based on the cloud-based IoT platform. In [22], Hao et al. suggest an architecture, which called Data Clouds, based on information-centric networking (ICN) to improve the accommodation of services for next generation of the Internet.

**7. IoT-A Architecture.** Another important architecture is the IoT-A European project [19] reference architecture, which is represented in Bassi et al. [23]. In [24], Pohls et al. inspire from the IoT-A architecture to design a framework for RERUM FP7 European Union project [16], which permit IoT applications to add privacy mechanisms and security in early design stages.

**8. S-IoT Architecture**. In [25], Atzori et al. merge IoT and social networks, and define the Social Internet of Things (S-IoT). They describe an architecture, which allows the integration of things in a social network and analyzes the components of the proposed network structure using simulations.

Nowadays, most of these architectures are implemented in industry or smart cities. Although these architectures are suitable for the moment, they will not be promising anymore in the sense of reliability and performance due to future challenges, and thus they will require being reexamined.

## III. NEW TECHNOLOGIES ENABLING IoT

In this section, we introduce several new technologies which are making an evolution in the future of the IoT applications. These technologies have not been originally designed for the IoT, but they can involve in the next generation IoT applications such as new services of smart cities and autonomous vehicles. To address such upcoming requirements, these technologies are to be embedded in the proposed architecture.

**1. Nano-chip**. Over the past two decades, Nano-Chip-based devices have found general applications in the analysis of biological and chemical samples. A tiny chip, which puts under the skin and by an electric field, reprogrammed cells could be an invention in the form we heal injured or aging tissue. Researchers say Nano-chip could cure injuries or with one touch, regrow organs. However, the usage of Nano-chip-based devices will not be limited to medical applications; for example, this technology could be used in military and home-automation applications which will cover the huge part of IoT applications.

**2. Millimeter Wave (mmWave)**. In last decade, the accessibility of frequency spectrum below 6 GHz bands have been coming down and the request for higher data rate is rising. The higher frequencies such as the millimeter wave (mmWave) which it frequency bands is above 24 GHz have been recommended as candidates for future 5G IoT applications because larger bandwidth could be considered to improve the capability and permit the users to use very high data rates for short range applications [26]. The frequency band of 24–28 GHz has exploited as one of the considered bands for 5G-IoT applications [27].

**3. Heterogeneous Networks (Het-Net).** Heterogeneous Networks (HetNet) has designed for satisfying the on-demand requirement of service-driving 5G-IoT. This novel-networking paradigm enables 5G-IoT to provide on-demand data transmission rate on request. In recent year, some 5G HetNet solutions have designed (See in [10]), which will deploy a massive number of resource-constrained devices.

**4. Direct Device to Device (D2D) Communication.** Direct Device-to-Device (D2D) Communication has designed as a new way for short-range data transmission, which will benefit the 5G-IoT with lower power consumption, better QoS for users and load balancing. The traditional Macro-cell Base Station (MBS) has considered supplying low power BS, However, D2D enable information transmissions between edge user equipment without of BS and serves as a "Cell Tier" in the 5G-IoT.

**5. Fifth Generation Wireless Systems (5G).** The fifth generation (5G) networks are becoming the main guide for the growth of IoT applications. The 5G can make important contributions to the next generation of IoT by connecting billions of intelligent things to generate actual future and massive IoT. At present, identification of IoT devices' capability is very difficult, because the heterogeneous domain of applications should satisfy the application's needs. According to the International Data Corporation (IDC) report, the services of the global 5G will support 70% of companies to spend $1.2 billion on the connectivity management solutions. Current IoT systems widely designed by specific application domain, such as BLE, ZigBee, other technologies, such as Wi-Fi, LP-WA networks (Lora, Sixgfox), and cellular communications (3GPP, LTE), etc. The IoT is rapidly developing with the new technology, especially the new application domain. Nowadays, IoT systems are improving the quality of lifestyles that involve the interconnection between smart home devices and smart environments. The Industry IoT (IIoT) is evolving many challenges, such as new requirements for product and solutions, and transforming business models [28]. the most popular communication techniques in the connectivity of IoT is 3GPP and LTE(4G) networks [29], which offer IoT systems with the reliabilities, timeless, robustness of connection, and with wide coverage, low deployment costs, high-security level, access to dedicated spectrum, and simplicity of management [30]. Yet, the existing cellular networks, for example; are not able to support the MTC communications but the 5G-IoT networks could provide it. In addition, 5G-IoT provides the fastest cellular network data rate with very low latency and improved coverage for MTC communication.

In recent years, various works on 5G-IoT have done [29]. The CISCO, Intel, Verizon, etc. have done wireless research projects on 5G, which adapted video quality to the requirements of the human eye [19]. The 5G-IoT provide real-time, reconfigurable, all-online, on-demand, and social experiences to IoT applications. The 5G-IoT architecture should be automatic and have to able end-to-end coordinated and intelligent and fast operation throughout each phase [9]. The 5G-IoT architectures will provide:

- Logically independent networks for applications' requirements;
- To reconstruct radio access network (RAN) engage cloud-based radio access network (Cloud RAN) for providing massive connections to multiple standards and implement on-demand deployment of RAN functions required by 5G.
- Simplifying the architecture of core network to design on-demand network functions configuration.

**6. Machine-Type Communication (MTC).** Machine-type communications (MTC) or machine-to-machine communications (M2M) represent automated data communications between the elemental infrastructure of data transport and devices. The data communications developed right between two MTC devices, or between an MTC device and a database [31]. MTC sets up a wide range of applications from a large deployment of autonomous devices to mission-critical services. Cellular systems (especially 5G) has considered as an important candidate to provide connectivity for MTC devices. MTC devices are more becoming an essential part of our lifestyle. The high data rate support and other salient features of MTC appear situation that 5G-Plus-HetNet has seen as a strong technology solution in 5G-IoT to the increasing data transfer demands from MTC devices [32].

**7. Wireless Software-Defined Networks (WSDN).** Wireless Software-Defined Networking (WSDN) is new technology that approaches to mobile cloud computing which aids network management and enables network configuration instead of improving network performance or network monitoring. The current networks require more flexibility and easy troubleshooting; to reach this subject, SDN breaks the vertical assimilation of traditional networks and through the centralized network control provide the flexibility to program the network. SDN is able to adapt the parameters of its network on the fly based on its operating conditions [33]. The 5G networks can be implemented through WSDN paradigm to provide faster and scalable 5G-IoT systems.

**8. Advanced Spectrum Sharing and Interference Management (Advanced SSIM).** The usable spectrum resource is limited and crowded. It may take years to redefine a spectrum band for the other usages, e.g. regularization or standardization of that are not easy. Spectrum efficiency is one of the important efficiency metrics in 5G communication networks. To enhance spectrum performance, advanced spectrum sharing techniques are usually used. Therefore, the 5G communication networks are expected to solve the issue with different methods [34].

**9. Mobile Edge Computing (MEC).** The Edge (Fog) computing is a distributed computing paradigm, which serves as a middle layer in between Cloud database and IoT devices/sensors. The Mobile Edge Computing (MEC) has proposed to describe the execution of services at the edge of the network, which aims to "provide an IT service environment and cloud-computing capabilities at the edge of the mobile network". MEC reference architectures and frameworks have the functional elements that support services such as location awareness, radio network information, and application execution. The advantages of expanding cloud services at the edge of mobile networks like 5G contain low latency, high bandwidth, and access to radio network information and location awareness. So that, it will be possible to optimize current mobile infrastructure services or implement new ones. MEC is another essential element in 5G-IoT, which will focus on two aspects:
- The analytics revolution, MEC, and 5G networks will be the core of the next generation of IoT;
- MEC in 5G-IoT will significantly increase computation related applications requiring massive processing, such as Virtual Reality (VR) or Augmented Reality (AR).

The before-mentioned deployment will reduce the costs, and provide a general management infrastructure for all virtualized services [35].

**10. Wireless Network Function Virtualization (WNFV).** Wireless Network Function Virtualization (WNFV) refer to network services and functions to wirelessly view network resources, such as databases, routers, links, and data, in a way that is separate from the general physical infrastructure, and to use these resources as service requirements as it needs. The WNFV separate a physical network into various virtual networks, therefore the devices can be reconfigured to organize various networks according to the requirements of applications. The WNFV as a supplementary to the 5G networks will enable the virtualization of the whole network functions for simplifying the deployment of 5G-IoT. The WNFV provides the scalable and flexible network for 5G-IoT applications, which will enable a customized network to create programmable networks for 5G-IoT applications. In fact, the WNFV will provide 5G-IoT applications online processing capability by optimizing the rate, ability, and coverage in the networks to suit the demands of applications. In addition, the WNFV will significantly improve the livability of radio access network (RAN). In the HetNet 5G-IoT, the network compression can aggregate 5G infrastructures with multi-RAT connectivity to befit the requirements of application services [36].

**11. Mobile Cloud Computing (MCC).** The cloud computing is an encouraging computing paradigm in the academia and industry, which refers to leverage virtualization technologies provide a variety of deployment models and service models, from public clouds to private ones, and from Infrastructure as a Service model to Software. Also, the cloud computing provides computing resources like processor, storage, and networks as a service. The benefits, such as efficient capability, on-demand self-service, accessibility, and scalability make cloud computing a computing resource opportunity for mobile devices. Due to the physical distance, cloud services cannot directly access local circumstantial information, like exact user location, local network circumstances, or situation of users' mobility behavior. For delay-sensitive applications, such as VR and AR, these requirements (such as mobility support low latency and context awareness) are expected. Mobile Cloud Computing (MCC) essentially focuses on the concept of 'mobile delegation': because of the available resource-constrained of mobile devices, the storage of bulk data and the execution of computationally intensive tasks should be delegated to remote entities. MCC refers to a paradigm, which leverages Cloud Computing resources to improve the performance of limited resources. Currently, many of services implemented in a centralized cloud or in the mobile devices themselves. There are several applications (such as AR and augmented interface (AI) applications) where the presence of a performance platform located at the environment of the mobile devices can present several advantages (such as lower latency and access to context information). Furthermore, as mobile devices are equipped with practical parts (such as sensors and high-resolution cameras) it is possible to generate collective sensing applications or novel crowdsourcing that attain the location information [35].

**12. Data analytics and Big Data.** One of the important aspects of a successful IoT application is Data Analytics. Companies are interested in pieces of information that brings value from massive data. The concept of Big Data is an abstract one, which is regularly dependent on the system's configuration (such as RAM, HDD space etc.) Recently, the value of Big data has been recognized, there are different opinions on Big Data definition. In fact, Big Data refers to the datasets that could not effectively be recognized, acquired, managed, and processed by common IT and software tools or hardware devices. Technologies such as Big Data and Data Analytics are changing the way we live and creating numerous new opportunities. The smart factory is one of the opportunities, which may contain hundreds of sensors generating massive amounts of data, Big Data. The needs of people could be more satisfaction by analysis of Big Data [4].

**13. Security and Forensics.** With the integration of 5G and the IoT, Security concerns such as privacy, access control, secure communication and secure storage of data are becoming important challenges in IoT Applications. Furthermore, every device that has built and every single data that has synchronized by an IoT application come under investigation. The large-scale distribution of the IoT devices and private nature of data that has accumulated and transferred by IoT nodes made security a primary challenge. It will not take long to see people suing each other for mistreating their smart gadgets, the thinking vehicles that have accident and criminals who put individual smart sensors at risk. The IoT has developed a haystack, which includes lots of helpful forensics artifacts while identification, collection, preservation, and reporting of pieces of evidence, such as an attack, would be challenging in this context [6].

The IoT will soon pervade all aspects of our life from managing our home temperature to thinking cars and smart management of the cities and the new technologies support this process. The researchers need a common language to use these technologies appropriately. Until now, these technologies all together are not compatible with current IoT architecture. Therefore, it is required to design a new architecture based on these technologies. In the next section, we show how these technologies contribute to the proposed IoT architecture.

## IV. PROPOSED ARCHITECTURE

In this section, we present an architecture that is suitable for the requirements of upcoming IoT applications and services. The new architecture is developed based on the technologies, which are explained in the last section, to provide a more sustainable, scalable, and mobile IoT ecosystem than existing IoT architectures. We propose an architecture based on 5G, which is called the 5G-IoT, with the following features: modular, efficient, agile, scalable, simple, and responsive to high demands.

The architecture consists of eight interconnected layers with two-way data-exchange capability as shown in Figure 1. The second layer and fifth layer consist of two and three sub-layers, respectively, and the security layer covers all other layers. These layers are selected to provide the best performance and maintain the modularity of the architecture simultaneously. The new technologies presented in Section III are embedded in the design of this new architecture to address the mentioned future challenges stated in Section I. The technologies with completely separate functionalities are embedded in different layers for ease of analysis, scalability, and modularity.

**L.1 Physical Device Layer.** This layer consists of wireless sensors, actuators, and controllers, which actually are the "things" of IoT. Physical devices are a common layer in all the architectures. In this layer, small size devices such as Nano-Chips are to be employed to increase computational processing power and to reduce power consumption. Nano-Chips are able to produce a high amount of initially processed data which is suitable for Big Data at data analytic layer (layer 7).

**L.2 Communication Layer.** This layer consist of two sub-layers. D2D Communication and Connectivity layers.

- **Direct Device To Device (D2D) Communication Sub-Layer.** Due to the increasing processing power and intelligence of physical devices (nodes), they contain their own identity and personality and generate their own data. To increase the efficiency and capabilities of the IoT systems, these devices should form a HetNet to communicate each other. Up-to-date communication protocols of the wireless sensor network (WSN) are exploited in this sub-layer. The nodes are able to do clustering or even select a leader (cluster head) for an appropriate networking. One of the most important technologies which improves this sub-layer is the mmWave. In addition, at this sub-layer, 5G is another optional technology which is able to enhance the D2D communication. The 5G has considered as an important candidate to provide connectivity for MTC devices. The high data rate support and other salient features of MTC make a situation that 5G-Plus-HetNet has been considered as a strong technology solution in the proposed 5G-IoT architecture.

- **Connectivity sub-layer.** In this sub-layer, devices are connected to communication centers, such as BSs. In addition, they send and analyze their data through the centers by the Intranet connection to the storage unit. At the moment, this sub-layer of the IoT has some specific problems: only limited number of device connections can be handled; in applications, such as autonomous vehicle, data exchange for a variety of data types are not applicable; high volume data can hardly be processed in real-time due to large communication latency. In near future, deployment of the 5G makes a great evolution at this sub-layer in the sense of reliability, performance, and agility. Another technology of this sub-layer is Advanced SSIM as explained in the last section. By this technology, the IoT devices attain the capability of choosing suitable spectrum (frequency bands) with sufficiently low interference. In fact, the SSIM techniques are beneficial to take the opportunities of spectrum sharing based on cognitive radios as well.

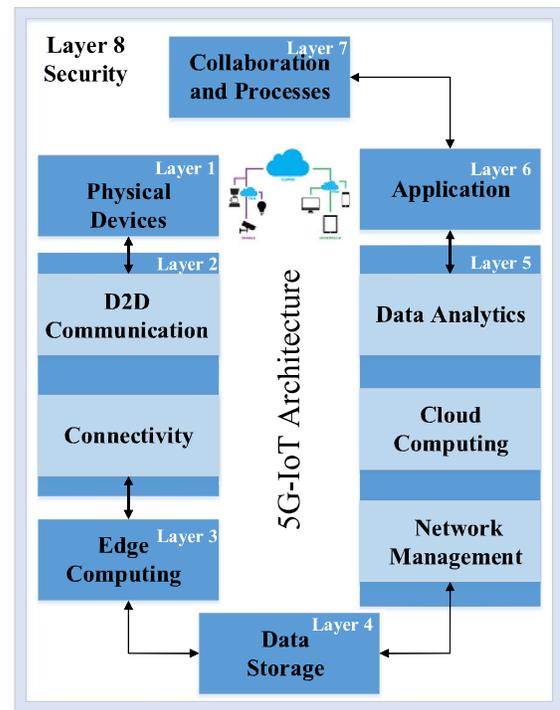

Figure 1. The proposed 5G-IoT architecture.

**L.3 Edge (Fog) Computing Layer.** At this layer, the data is edge processed by nodes or their leaders in order to make decisions at the Edge level. With the introduction of 5G technology and the rise of mobile devices (such as smartphones), MEC technology will be more powerful to overcome the challenges, and will significantly contribute at this layer.

**L.4 Data Storage layer.** This layer contains data storage units in which the information obtained from edge processing of the physical devices are stored as well as raw data. This layer requires special protection in terms of security, and also should be responsive to the huge data volume and traffic of future applications.

**L.5 Management Service layer.** This layer consist of three sub-layers as follows:

- **Network Management Sub-Layer.** Network management involves changing the type of communication between devices and data centers. The most important technology involved in this sub-layer is WNFV. The WNFV is able to update topology of the network and type of communication protocols, such as 5G-IoT or ZigBee, simultaneously to improve the quality of the IoT structure. Another useful technology at this sub-layer is WSDN. The WSDN manages the IoT network and enables network reconfiguration instead of traditional network monitoring for performance enhancement.

- **Cloud Computing Sub-Layer.** In this sub-layer, data and information from the edge computing are (re)processed in the cloud so that final processed information can be derived. By implementation of 5G technology, the mobile devices are capable of performing this type of computing, which is referred as MCC, in real-time. Hence, the processing operations will be distributed among mobile devices in parallel to make the IoT system more efficient, sustainable, scalable, and faster.

- **Data Analytics Sub-Layer.** In this sub-layer, new methods of data analytics [4] are employed to produce value (manipulatable information) from raw data. Any improvement in Big Data algorithms will enhance the data processing at this sub-layer. In fact, the role of this sub-layer is dominant in near future when the amount of collected information is increased due to the integration of 5G and the IoT.

**L.6 Application Layer.** In this layer, Software interacts with previous layers and data, which is at rest, so it is not necessary to operate at speed of the network. Applications are able to revolutionize vertical markets and business needs by control applications, Vertical and Mobile Applications and Business Intelligence and Analytics. In fact, Applications layer let the business people do the right thing at the right time by the right data.

**L.7 Collaboration and Processes Layer.** The IoT system and the information arrives from the previous layers are not useful unless it produces an act. People are empowered by applications performing business logic. People use applications and associated data for their particular needs. Sometimes, multiple individuals use the same application for different purposes. In fact, individuals must be able to collaborate and communicate to make the IoT serviceable.

**L.8 Security Layer.** Like many architectures [23], [24] this layer is considered to be a separate layer. In fact, this layer covers and protects all previous layers but each section (the intersection of this layer with another layer) has its own functionality. The security layer of the proposed architecture entails various terms of security features including data encryption, user authentication, network access control and cloud security [6]. In addition, security layer also prevents and anticipates the dangers and cyber-attacks, including the forensics to detect the type of attack and fail them.

## V. COMPARISONS

In Section II, the most important IoT architectures were introduced. In this section, we compare those architectures with the proposed architecture. As we mentioned, the next generation IoT applications and their services, such as smart factory and smart city, require special attributes, as follows: Support of Variety of data types, Support a high number of customers and demands, Agility, Flexibility, Robustness of connection, Low Latency Reliable Communication

The current IoT architectures with their communication and networking technologies suffer from providing the above requirements [16]. However, the architectures based on 5G communication technology are able to satisfy the above requirements and provide the following features: Simplicity of management, Reliability, Reconfigurability, High security, Easy and fast troubleshooting, Wide coverage, Low deployment cost. As far as we know, the number of IoT architecture, which is based on 5G Communication, is limited to recent publications [10]. Moreover, most of them only concentrate on 5G technology without enough attention to other upcoming technologies. In fact, with the integration of 5G and the IoT, many other technologies such as MTC and WNFV could take a part in next-generation IoT architectures. The appropriate combination of these technologies can create a more comprehensive structure, which meets the listed requirements of next-generation IoT applications. This approach is not addressed by the introduced architectures in Section II as shown in Table 1.

Table 1. IoT Architectures Comparisons.

|  | Low Latency | Robustness of connection | Support of different data types | Reconfigurability | Wide coverage |
|---|---|---|---|---|---|
| SDN Architecture | ✗ | ✗ | ✗ | ✓ | ✗ |
| QoS Architecture | ✗ | ✓ | ✗ | ✓ | ✗ |
| Three level Architecture | ✗ | ✗ | ✗ | ✗ | ✗ |
| SoA Based Architecture | ✗ | ✓ | ✗ | ✓ | ✗ |
| S-IoT Based Architecture | ✗ | ✓ | ✗ | ✗ | ✗ |
| IoT-A Architecture | ✗ | ✓ | ✗ | ✗ | ✓ |
| The Proposed Architecture | ✓ | ✓ | ✓ | ✓ | ✓ |

## VI. CONCLUSIONS

In this paper, a novel architecture has proposed which considered future requirements of new applications and their generated data. Also, it presents a new model that consists of Nano-chip, millimeter Wave (mmWave), Heterogeneous Networks (HetNet), device-to-device (D2D) communication, 5G-IoT, Machine-Type Communication (MTC), Wireless Network Function virtualization (WNFV), Wireless Software Defined Networks (WSDN), Advanced Spectrum Sharing and Interference Management (Advanced SSIM), Mobile Edge Computing (MEC), Mobile Cloud Computing (MCC), Data Analytics and Big Data. Based on these technologies, a novel architecture has been proposed. The architecture is modular, efficient, scalable, reliable, simple, and it is able to support high application demands. Indeed, the proposed architecture is able to meet the needs of the next generation IoT applications and assist IoT experts to design more efficient and scalable IoT systems.